\documentclass{elsart}

 \usepackage{graphicx}

\begin{document}
PCP-2 /ISS 2001

\begin{frontmatter}



\title{Critical Temperature $T_{\rm c}$ versus Charging Energy $E_{\rm c}$
in MgB$_{2}$
and C$_{60}/$CHBr${_3}$}


\author[okayama1]{Chikao Kawabata\corauthref{cor1}},
\ead{kawabatact@nifty.com}
\author[okayama2]{Nobuhiko Hayashi},
and
\author[okayama3]{Fumihisa Ono}

\address[okayama1]{Division of Liberal Arts and Sciences, Okayama University,
Okayama 700-8530, Japan}
\address[okayama2]{Computer Center, Okayama University,
Okayama 700-8530, Japan}
\address[okayama3]{Department of Physics, Okayama University,
Okayama 700-8530, Japan}

\corauth[cor1]{Corresponding author.\\
Postal address: Faculty of Environmental Science and Technology,
Okayama University, 3-1-1 Tsushima-naka, Okayama 700-8530, Japan\\
Fax: +81-86-273-6750}

\begin{abstract}
   The boride compounds MB$_x$ related to the magnesium-boron stacking layered
material MgB$_2$
are discussed in terms of
the B-B layers in the borides analogous to the Cu-O
ones in the cuprates.
   We propose a possibility of
superconducting materials
which exhibit higher critical temperature $T_{\rm c}$ than 39 K of MgB$_{2}$.
   We point out a role of
interstitial ionic atoms M (e.g., Mg in MgB$_{2}$)
as capacitors,
which reduce the condensation-energy loss due to the charging energy
$E_{\rm c}$ between the B-B layers.
   In the viewpoint of the present model,
   the recently discovered 117-Kelvin superconductor C$_{60}/$CHBr$_3$
is also discussed in terms of the intercalation molecules CHBr$_3$
as possible capacitors among the superconducting grains of C$_{60}$ molecules.
\end{abstract}

\end{frontmatter}

\section{Introduction}
\label{intro}
   Since Akimitsu and co-workers presented a remarkably high
temperature ($\simeq$ 39 K)
superconducting magnesium diboride MgB$_{2}$
in the beginning of 2001~\cite{nagamatsu01,buzea01},
much attention
has been focused on this diboride material
and some related boride compounds
\cite{buzea01,felner01,young01,materials}.
   The material MgB$_{2}$
takes the layered structure of B-Mg-B stacking and has a coupling between
2-dimensional (2D)
B-B layers~\cite{nagamatsu01,buzea01,bianconi01}.
   According to
several band calculations
~\cite{kortus01,an01,satta01,bands},
in the MgB$_{2}$-type crystal structure
there exist a strong covalent B-B bonding
and an ionic B-Mg bonding.
   The strong covalent B-B bonding and the presence of the ionic Mg atoms
lead to hole-like
cylindrical Fermi surfaces in MgB$_{2}$~\cite{kortus01,an01},
and the holes are conducted in each 2D B-B layer.
   It is observed that
the 2D B-B layers,
like the Cu-O layers in high-$T_{\rm c}$ cuprates, play an important role
in the electrical transport properties~\cite{jin01}.
The superconductor
MgB$_{2}$ can then be beautiful realization of the essential physics of
superconductivity in cuprates, without the complications of
Cu $d_{x^2-y^2}$ orbitals~\cite{hirsch01}.

   In a previous paper~\cite{kawabata01},
on the basis of a theoretical model
of the B-B layers analogous to the Cu-O
layers~\cite{kawabata94a,kawabata95a,kawabata95b,kawabata94b,kawabata96},
we investigated the ionic diboride compounds (M$^{2+}$)(B$^{-})_2$,
$\bigl($e.g., (Mg$^{2+}$)(B$^{-})_2$ in the case of MgB$_2$$\bigr)$,
and discussed their superconducting critical temperature $T_{\rm c}$.
   We made investigation into $T_{\rm c}$
assuming a simple linear relation
between $T_{\rm c}$ and the inter-layer charging energy
$E_{\rm c}$~\cite{kawabata01}.
   In the present paper,
instead of assuming the linear relation,
we will analyze $T_{\rm c}$ and $E_{\rm c}$
in more detail
on the basis of an original formula for $T_{\rm c}$ presened
in Refs.~\cite{kawabata94a,kawabata95a,kawabata95b},
and also mention general boride compounds
MB$_x$ in which the interstitial ionic atoms M
between B-B layers
are not restricted to M$^{2+}$.
   The purpose of this paper is to grope for various possible
boride compounds with $T_{\rm c}$ over 39 K and
to give a hint as to new material synthesis.
   In the viewpoint of our theoretical model,
the recently discovered 117-Kelvin superconductor
C$_{60}/$CHBr$_3$~\cite{Schoen01}
is also discussed.

\section{Physical Model}
\label{model}
   We model the boride compounds MB$_x$
(M$=$ Mg, Be, etc.)
as quantum-capacitive
Josephson junction arrays (JJA)
with weakly coupled superconducting grains
on a 3D lattice,
following Kawabata-Shenoy-Bishop (KSB)
theory~\cite{kawabata94a,kawabata95a,kawabata95b,kawabata94b,kawabata96},
originally proposed to the high-$T_{\rm c}$ cuprates
with 2D Cu-O layers and a coupling between the Cu-O
layers.
   In the present JJA model,
the Josephson coupling energy $E_{\rm J}$ via
the Cooper pair tunneling within and between the B-B layers enhances
the critical temperature $T_{\rm c}$,
while the charging energy $E_{\rm c}$
between the B-B layers
depresses $T_{\rm c}$.
   The origin of $E_{\rm c}$ is that,
when the Cooper pairs tunnel
between the layers, local inter-layer charge unbalance is induced
between the superconducting grains in JJA.
   This charge unbalance gives rise to the electric fields
between the grains and it costs the electromagnetic energies
and destroys the phase coherence of
superconductivity~\cite{kawabata94a,kawabata95a,kawabata95b,simanek}.
   Here, an important point for our analysis is that
the interstitial ionic atoms, M,
located between the B-B layers play a role of
capacitors,
reducing the charging energy $E_{\rm c}$
by the electric polarization of the ionic atom.

   We estimate $T_{\rm c}$ which is a function
of
$E_{\rm c}$.
   According to the KSB theory~\cite{kawabata94a,kawabata95a,kawabata95b},
in the present JJA model $T_{\rm c}$ is expressed, with
an unknown function $f$, as
\begin{eqnarray}
\frac{E_{\rm J}}{k_{\rm B}T_{\rm c}}
=0.454f(\gamma_0^{-2},E_{\rm c}/E_{\rm J}),
\label{eq:Tc-KSB-1}
\end{eqnarray}
and for zero charging energy $E_{\rm c}=0$ it must reduce to
a known equation
\begin{eqnarray}
\frac{E_{\rm J}}{k_{\rm B}T_{\rm c}}
=0.454f(\gamma_0^{-2},0),
\label{eq:Tc-KSB-2}
\end{eqnarray}
which is the dimensionless critical coupling~\cite{shenoy94,shenoy95}
in terms of the coupling anisotropy $\gamma_0^{-2}$
in the classical anisotropic 3D XY/JJA system
$\bigl($$\gamma_0^{-2}=E_{\rm J \perp}/E_{\rm J \parallel}$,
where $E_{\rm J \perp}$ is the inter-layer Josephson coupling energy,
$E_{\rm J \parallel}$ the intra-layer one, and
$E_{\rm J}=\sqrt{E_{\rm J \perp} E_{\rm J \parallel}}$$\bigr)$.
   In the 3D limit $\gamma_0^{-2} \rightarrow 1$,
the present JJA model may be applied to the superconductor
C$_{60}/$CHBr$_3$,
where the coupled superconducting grains on a 3D lattice
in the 3D JJA are composed of C$_{60}$ molecules.
   We discuss this superconductor in Sec.\ \ref{C60}.

   Expanding $f$ with respect to $E_{\rm c}/E_{\rm J}$, we obtain
from Eq.\ (\ref{eq:Tc-KSB-1})
the expression for $T_{\rm c}$ as~\cite{kawabata94a,kawabata95a,kawabata95b}
\begin{eqnarray}
k_{\rm B}T_{\rm c}
\sim \frac{E_{\rm J}}{f_0+f'_0 E_{\rm c}/E_{\rm J}},
\label{eq:Tc-KSB-3}
\end{eqnarray}
where $f_0=f(\gamma_0^{-2},0)$.
   In this paper, we rewrite Eq.\ (\ref{eq:Tc-KSB-3}) as
\begin{eqnarray}
T_{\rm c} & = & \frac{T_{\rm c}^{max}}{1+ A E_{\rm c}} \nonumber \\
          & \equiv & F(E_{\rm c}),
\label{eq:Tc-KSB-4}
\end{eqnarray}
with the phenomenological parameters
$T_{\rm c}^{max}$ and $A$,
which should be determined empirically.
We use Eq.\ (\ref{eq:Tc-KSB-4}) to fit data and estimate $T_{\rm c}$
for some borides
in the next section.

\section{Estimation of $E_{\rm c}$ and $T_{\rm c}$ in the boride compounds}
\label{estimation}
   Referring to Kittel's textbook and
Shockley's
tables~\cite{kittel,tessman53,pauling27,shanon}
for the ionic radius $d$ and electronic
polarizability $\varepsilon$ of the ionic atoms M,
we have obtained the charging energy
$E_{\rm c} \sim d / \varepsilon$
between the B-B layers as in Table~\ref{table:Ec}, following
the KSB theory
~\cite{kawabata94a,kawabata95a,kawabata95b,kawabata94b,kawabata96}.

\begin{figure}
\includegraphics[width=14cm]{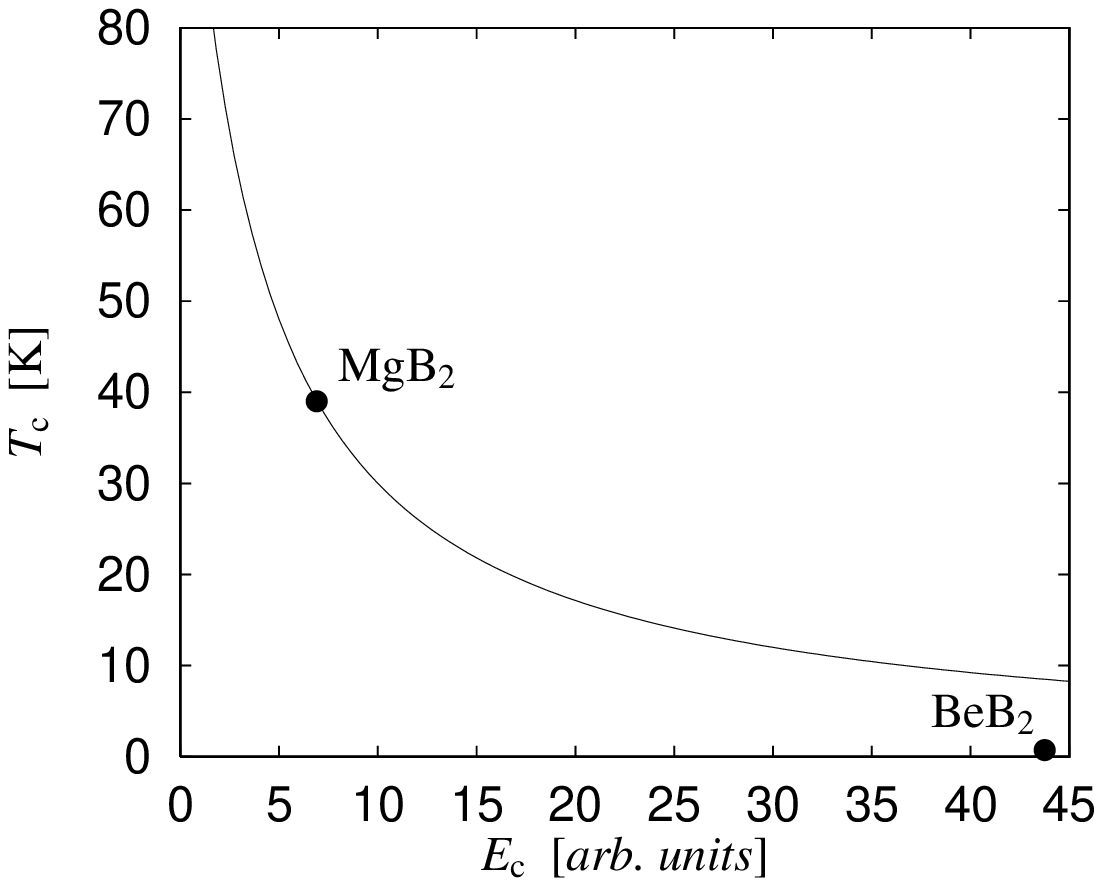}
\vspace{1cm}
\caption{
   The plot of
the critical temperature $T_{\rm c}$
versus charging energy $E_{\rm c}$ (see text).
   Points are for MgB$_2$ ($T_{\rm c} = 39$ K~\cite{nagamatsu01})
and BeB$_2$ ($T_{\rm c} \simeq 0$ K ($<5$ K)~\cite{felner01}).
   The values of $E_{\rm c}$ are of Table~\ref{table:Ec}
for Mg$^{2+}$ and Be$^{2+}$.
}
\label{fig:tc}
\end{figure}

   We discuss first
the ionic compounds M$^{2+}$(B$^{-}$)$_2$
expected to have nearly the same electronic structure~\cite{satta01}
as MgB$_{2}$ (Mg$^{2+}$(B$^{-}$)$_2$)
and to have almost the same Josephson coupling energy $E_{\rm J}$
$\bigl[$i.e., the same ``bare $T_{\rm c}$''  ($T_{\rm c}^{max}$)
without any influence of
the M-atom-dependent charging energy $E_{\rm c}$,
and the same value of $A$ in Eq.\ (\ref{eq:Tc-KSB-4})$\bigr]$.
   By hypothesizing the same $E_{\rm J}$ for
any M$^{2+}$(B$^{-}$)$_2$ compounds,
we estimate critical temperatures
$T_{\rm c}$ ($=F(E_{\rm c})$).
   In Fig.~\ref{fig:tc}
we show two points,
as experimentally known data,
for BeB$_{2}$ ($T_{\rm c} \simeq 0$ K ($<5$ K)~\cite{felner01},
$E_{\rm c}=43.75$)
and MgB$_{2}$ ($T_{\rm c}=$39 K~\cite{nagamatsu01}, $E_{\rm c}=6.91$),
where $E_{\rm c}$ ($\sim d/\varepsilon$) are of Table~\ref{table:Ec}
and in arbitrary units.
   From Eq.\ (\ref{eq:Tc-KSB-4}) and the values of
$T_{\rm c}$ and $E_{\rm c}$ for MgB$_{2}$,
we obtain an equation $T_{\rm c}^{max}=39 \times (1+6.91 A)$.
   From this equation and Eq.\ (\ref{eq:Tc-KSB-4}) with
$T_{\rm c} <5$ K and $E_{\rm c}=43.75$ for BeB$_{2}$,
we obtain $A < 0$, which is contrary to the condition
that the charging energy $E_{\rm c}$ decreases $T_{\rm c}$
in Eq.\ (\ref{eq:Tc-KSB-4}).
   It should be remedied by taking
higher order terms of $E_{\rm c}/E_{\rm J}$
when obtaining Eq.\ (\ref{eq:Tc-KSB-3}).
   Instead, we plot the $T_{\rm c}$ vs.\ $E_{\rm c}$ curve
in Fig.~\ref{fig:tc}
on the basis of Eq.\ (\ref{eq:Tc-KSB-4}), assuming
$T_{\rm c}^{max}=120$ K as representative.
   In this case $T_{\rm c}($BeB$_{2})=8.6$ K at $E_{\rm c}=43.75$.
   Even if $T_{\rm c}^{max}$ is increased further,
$T_{\rm c}$ at $E_{\rm c}=43.75$
only decreases a little.
   On the other hand, when $T_{\rm c}^{max}$ is decreased,
$T_{\rm c}$ at $E_{\rm c}=43.75$ is increased and
deviates more from the experimental result~\cite{felner01}
$T_{\rm c}($BeB$_{2}) <5$ K.
   We then believe that
the $T_{\rm c}$ vs.\ $E_{\rm c}$ curve
drawn in Fig.~\ref{fig:tc} is not so unreasonable.
   It is expected that
the diborides, which contain the ionic atoms M with
the electric polarizability $\varepsilon$
and have the corresponding $E_{\rm c}$ in Table~\ref{table:Ec},
are distributed on the $T_{\rm c}$ vs.\ $E_{\rm c}$ graph
along the curve drawn in Fig.~\ref{fig:tc}.

   In Ref.~\cite{young01}, $T_{\rm c}=0.72$ K
is observed in BeB$_{2.75}$, which has the crystal structure
different from MgB$_{2}$.
   It has also been suggested that nonstoichiometry may be an important point
in the superconductivity of some boride compounds~\cite{buzea01}.
   It is then interesting to synthesize various borides MB$_x$
and try to measure their $T_{\rm c}$.
   We give in Table~\ref{table:Ec}
various atoms M for which the polarizability has been
obtained~\cite{tessman53,pauling27}.
   The present theory can be applied to
the borides MB$_x$ which have the B-M-B stacking structure.
   In this case the values of the Josephson coupling energy $E_{\rm J}$ and
the coupling anisotropy $\gamma_0^{-2}$
$\bigl($and thus, the corresponding $T_{\rm c}^{max}$ and $A$
in Eq.\ (\ref{eq:Tc-KSB-4})$\bigr)$
would be different in general for each compound.
   It is, however, expected that the compounds with smaller $E_{\rm c}$
is adventageous to the superconductivity,
which can be a guide in the synthesis of new materials.
   The thin films of MB$_x$ with the B-M-B stacking structure
could be synthesized by, for example, the epitaxial methods.

\section{Discussion of superconductor C$_{60}/$CHBr$_3$}
\label{C60}
   Quite recentry,
Sch\"on {\it et al.}~\cite{Schoen01} found a
new superconducting system C$_{60}/$CHBr$_3$
which exhibits very high critical temperature
$T_{\rm c}=117$ K.
   It is  a C$_{60}$ superconductor intercalated with CHBr$_3$ molecules.
   This $T_{\rm c}$ is about two times larger than
52 K~\cite{Schoen00} of C$_{60}$ without intercalations.
   The critical temperatures of C$_{60}$ superconductors
increase with the lattice constants $a$;
C$_{60}$ ($T_{\rm c}=52$ K, $a=14.15$ \AA),
C$_{60}/$CHCl$_3$ ($T_{\rm c} \sim 70$ K, $a=14.29$ \AA), and
C$_{60}/$CHBr$_3$ ($T_{\rm c}=117$ K, $a=14.45$ \AA)~\cite{Schoen01}.
   One of the promising course of attaining higher $T_{\rm c}$
is certainly to search for new spacer molecules
which expand the separation between C$_{60}$ molecules~\cite{Schoen01}.
   However,
it seems to be open to
further discussion
whether the only 2 \% expansion of the lattice constant
is sufficient for the increase of the electronic density of states
which corresponds to rather drastic rise of $T_{\rm c}$
from 52 to 117 K.

   In the viewpoint of our theory,
we would like to point out another possibility that
the intercalation molecules would play a role of the capacitors
in the C$_{60}$ superconductors
as discussed for the atoms M in the boride compounds MB$_x$.
   By the electric polarization of the neutral molecules,
the intercalation molecules such as CHBr$_3$
may reduce the charging energy $E_{\rm c}$ between
the C$_{60}$ molecules which constitute
a 3D JJA as superconducting grains
on a 3D lattice.
   The smaller the charging energy $E_{\rm c}$ is,
the higher the critical tempereture $T_{\rm c}$ is,
as discussed in the preceding sections.
   According to
experimental data listed in Table 11 of Ref.~\cite{duijnen98},
the molecular polarizability of CHBr$_3$ is indeed
larger than that of CHCl$_3$,
which is in accord with the present scenario that
the intercalation molecules act as capacitors which reduce $E_{\rm c}$.

   We can estimate $T_{\rm c}$ of other
intercalated C$_{60}$ superconductors
as follows.
   The experimental values of the molecular polarizability
are $\varepsilon$(CHCl$_3)=57.56$ au and
$\varepsilon$(CHBr$_3)=79.9$ au~\cite{duijnen98}.
   Assuming an approximately constant value of the lattice constant $a$,
the charging energies $E_{\rm c} \sim a/\varepsilon \sim \varepsilon^{-1}$
in arbitrary units are obtaind as
$E_{\rm c}($C$_{60}/$CHCl$_3)=0.0174$ and
$E_{\rm c}($C$_{60}/$CHBr$_3)=0.0125$.
On the $T_{\rm c}$ vs.\ $E_{\rm c}$ graph,
extrapolating the straight line
which goes through two points
$(T_{\rm c},E_{\rm c})=(70,0.0174)$ for C$_{60}/$CHCl$_3$
and $(117,0.0125)$ for C$_{60}/$CHBr$_3$,
namely, $T_{\rm c}=-9592 E_{\rm c} + 237$,
we predict $T_{\rm c} \simeq 158$ K (C$_{60}/$CHI$_3$),
127 K (C$_{60}/$CH$_{2}$I$_{2}$),
137 K (C$_{60}/$C$_6$H$_{4}$Cl$_{2}$),
127 K (C$_{60}/$C$_6$H$_{5}$NO$_{2}$),
121 K (C$_{60}/$C$_6$H$_{5}$Cl),
and
175 K (C$_{60}/$C$_{12}$H$_{26}$).
Here, we use $\varepsilon=121.74$ au (CHI$_3$),
87.05 au (CH$_{2}$I$_{2}$),
95.83 au (C$_6$H$_{4}$Cl$_{2}$),
87.19 au (C$_6$H$_{5}$NO$_{2}$),
82.67 au (C$_{60}/$C$_6$H$_{5}$Cl),
and
153.86 au (C$_{12}$H$_{26}$)~\cite{duijnen98}.
   Depending on actual lattice constants,
other molecules with relatively large polarizability
listed in Ref.~\cite{duijnen98}
may also be interesting as candidates of the intercalation molecules.


\section{Conclusions}
\label{conclusion}
   We indicated that, if the boride compounds are
synthesized by using ionic atoms with larger polarizability
$\varepsilon$
or smaller diameter
instead of the Mg atoms, the critical temperatures $T_{\rm c}$
will become higher
than 39 K of MgB$_2$.
  We also pointed out that the recently descovered superconductor
C$_{60}/$CHBr$_3$ will owe
its very high critical temperature
to the intercalation molecules CHBr$_3$
{\it as electoric capacitors} which reduce
the charging energy $E_{\rm c}$ $(\sim a/\varepsilon)$
among the superconducting grains of C$_{60}$ molecules,
rather than as simple spacer molecules
which expand the separation $a$ between C$_{60}$ molecules.
   We hope that
the present proposals are a hint or
an encouragement to future synthesis of new materials.

\begin{ack}
   One of the authors (C.K.) would like to thank
Professor S.\ R.\ Shenoy,
Dr.\ A.\ R.\ Bishop
and
Dr.\ N.\ L.\ Saini
for discussions and encouragement
in the initial stage of this study.
\end{ack}




\newpage
\
\thispagestyle{empty}

\newpage

\thispagestyle{empty}

\begin{table}
\caption{
   Ionic radius $d$
\cite{kittel}
ionic electric polarizability $\varepsilon$
\cite{tessman53}
and charging energy $E_{\rm c} \sim d/\varepsilon$.
   The data of $\varepsilon$ in brackets are
of Ref.\
\cite{pauling27}
originally
and the values of $d$ for Cu are of Ref.\
\cite{shanon}
}
\vspace{-2.1mm}
\begin{tabular}{lllll}
Ionic &
Ionic &
Ionic &
Charging &
Critical \\
atom &  radius~\cite{kittel,shanon} &
 electronic &
 coupling &
temperature \\

 & &
polarizability~\cite{tessman53,pauling27} &
energy &
 \\
M    &    $d$ [10$^{-8}$cm]  & $\varepsilon$ [10$^{-24}$cm$^3$]  &
$d/\varepsilon$ ($E_{\rm c}$)  &   $T_{\rm c}$ [K]\\
\hline
Be$^{2+}$ &    0.35 &    (0.008)      &        (43.75)    &
$\simeq 0$ ($<5$) ~\cite{felner01} \\
 & & & &   0.72 ~\cite{young01} \\
Mg$^{2+}$ &    0.65 &    (0.094)      &        (6.91)     &
     39 ~\cite{nagamatsu01}\\
Ca$^{2+}$ &    0.99 &    1.1 (0.47)  &      0.9  (2.11) \\
Sr$^{2+}$ &    1.13 &    1.6 (0.86)  &      0.71 (1.31) \\
Ba$^{2+}$ &    1.35 &    2.5 (1.55)  &      0.54 (0.87) \\
Zn$^{2+}$ &    0.74 &    0.8      &         0.93 \\
Cd$^{2+}$ &    0.97 &    1.8      &         0.54 \\
Cu$^{2+}$ &    0.73 &    0.2      &         3.65 \\
Pb$^{2+}$ &    0.84 &    4.9      &         0.17 \\
Li$^{+}$ &    0.68 &    0.03 (0.029)    &   22.67 (23.45) \\
Na$^{+}$ &    0.97 &    0.41 (0.179)    &   2.37  (5.42) \\
K$^{+}$  &    1.33 &    1.33 (0.83)     &   1     (1.6) \\
Rb$^{+}$ &    1.48 &    1.98 (1.4)      &   0.75  (1.06) \\
Cs$^{+}$ &    1.67 &    3.34 (2.42)     &   0.5   (0.69) \\
Cu$^{+}$ &    0.77 &    1.6             &   0.48 \\
Ag$^{+}$ &    1.26 &    2.4             &   0.53 \\
Tl$^{+}$ &    0.95 &    5.2             &   0.18 \\
Al$^{3+}$ &    0.5  &    (0.052)         &  (9.62) \\
Sc$^{3+}$ &    0.81 &    (0.286)         &  (2.83) \\
Y$^{3+}$  &    0.93 &    (0.55)          &  (1.69) \\
La$^{3+}$ &    1.15 &    (1.04)          &  (1.11) \\
C$^{4+}$   &    0.15 &    (0.0013)        & (115.38) \\
Si$^{4+}$  &    0.41 &    (0.0165)        & (24.85) \\
Ti$^{4+}$  &    0.68 &    (0.185)         & (3.68) \\
Ge$^{4+}$  &    0.53 &    1               & 0.53 \\
Zr$^{4+}$  &    0.8  &    (0.31)          & (2.58) \\
Sn$^{4+}$  &    0.71 &    3.4             & 0.21 \\
Ce$^{4+}$  &    1.01 &    (0.73)          & (1.38)
%
\end{tabular}
\label{table:Ec}
\end{table}

\end{document}